\documentclass[a4paper,10pt]{article}
\usepackage[utf8x]{inputenc}
\usepackage{amssymb}
\title{Noncommutativity and Non-anticommutativity in Perturbative Quantum 
Gravity}
\author{Mir Faizal\\ 
Mathematical Institute,  University of Oxford,\\ Oxford, OX1 3LB, 
 United Kingdom}

\begin{document}

\maketitle

\begin{abstract}
In this paper we will study 
 perturbative quantum gravity on supermanifolds  with
both noncommutative and non-anticommutative 
 coordinates. 
We shall first   analyses the BRST and the anti-BRST symmetries of this 
 theory. Then 
 we will also analyze the effect of 
shifting all the fields of this theory in background 
field method. We will construct a 
 Lagrangian density which apart from being  invariant 
under  the extended  BRST transformations is also invariant under on-shell
extended anti-BRST transformations. This will be done by using 
  the  Batalin-Vilkovisky   (BV) formalism. 
Finally, we will show that the sum of  
the gauge-fixing term and the ghost term for this 
theory  
 can be elegantly written  down in superspace with 
 two Grassmann parameters. 
\end{abstract}


\section{Introduction}
Noncommutative gauge theory  first appeared due 
to $NS$ tensor backgrounds \cite{2pas}-\cite{3p}. 
Thus, many models based on the noncommutative gauge theories, 
including noncommutative standard model has been studied \cite{1ab}-\cite{2ab}.
 However, due to AdS/CFT correspondence  a superconformal
 noncommutative gauge theory on the boundary of
 anti-de Sitter spacetime is dual
 to  the noncommutative gravity in the bulk of that anti-de Sitter 
 spacetime \cite{ads1}-\cite{ads2}.
The $RR$ background in string theory leads to a
 non-anticommutative gauge theory \cite{ov}-\cite{se}. Thus, 
many models based on 
non-anticommutative  gauge theory have also been studied \cite{ss1}-\cite{ss2}. 
Similarly, in analogy with noncommutative case, a superconformal
non-anticommutative gauge theory on the boundary of anti-de Sitter 
 spacetime
 will be dual to non-anticommutative  gravity in the bulk of that
 anti-de Sitter  spacetime. 
Hence, it will
 be intresting to study both commutative and non-anticommutative
 gravity in anti-de Sitter  spacetime. 
 Non-anticommutative gravity has already been studied 
in flat spacetime 
\cite{flat1}-\cite{flat}. This work can provide the framework for 
analysing non-anticommutative gravity in anti-de Sitter  
spacetime. In fact,   
it is possible to analyse gravity with
 both noncommutative and non-anticommutative coordinates 
in the supermanifold formulism. 
Quantum field theory on spacetime with 
both noncommutative and non-anticommutative coordinates
has already  been analyzed in the supermanifold formulism \cite{k0}.
 Unlike,  noncommutative quantum field theory,
 this  model of quantum field theory
 differs from the conventional 
 quantum field theory even in absence of interactions.  
This is because even the 
 Feynman propagator get modified due to the non-anticommutativity 
of the spacetime. 

It will be interesting to generalise
 this work to anti-de Sitter  spacetime and use it to study the model of
 gravity that is dual to both the gauge theories 
on  $NS$ and $RR$ backgrounds. However, 
we will not do that in this paper. In this paper we will rather study
 some more formal aspects relating to gauge fixing of the 
noncommutative and non-anticommutative gravity in flat spacetime.
 This work along with the previous work done in this 
direction can provide the bases for analysing noncommutative 
and non-anticommutative gravity
 in anti-de Sitter  spacetime which intern can be used to study an
 interesting 
example of AdS/CFT correspondence. 

It may be remarked that non-anticommutativity 
in gravity has also occurs in complex spacetime \cite{299p}-\cite{229p1}. 
These 
models of gravity on a complex spacetime were initially 
studied as attempts to unify gravity with electromagnetism \cite{k4}. 
However, they are now studied due to their relevance in string theory
\cite{k6}-\cite{k8}.  
Apart from this many models of quantum gravity have  suggested that
 the Plancks length might act like a minimum length scale for spacetime and 
this has led to the development of a modification of the General 
relativity called the Rainbow gravity \cite{gr1}-\cite{gr2}. 
Such a minimum 
length occurs naturally
in a  
spacetime with a combination of  noncommutative 
and non-anticommutative coordinates \cite{flat}.  
Another advantage of using noncommutative
 or non-anticommutative coordinates is that 
it gives spacetime a fuzzy structure \cite{fuzzy}-\cite{fuzzy1}. 
This fuzziness   
 of the spacetime can be used to solve the problems related to
the occurrence of the 
singularity in a black hole \cite{bh1}-\cite{bh2}. 

All the degrees of freedom of perturbative quantum gravity on 
supermanifolds  are not physical. Thus, before quantizing 
this theory we have to fix a gauge. This can be done at by adding a gauge fixing term and a ghost term to the original 
classical Lagrangian density. The effective Lagrangian density obtained by the sum of the gauge fixing term and the ghost term 
with the original classical Lagrangian density is invariant under BRST transformations. 
It is well know that any theory which is invariant under BRST transformations is also invariant under anti-BRST transformations. 
In these anti-BRST transformations, the role of ghosts and anti-ghosts is almost reversed. 

Furthermore, in background field method all the field of the theory are shifted.  
The invariance of the theory under the original BRST and the original anti-BRST transformations along with these shift transformations 
  can be analyzed in the  Batalin-Vilkovisky (BV) formalism  
\cite{1}-\cite{3f}. 
The BRST symmetry and shift 
symmetry for perturbative quantum gravity on non-anticommutative
 spacetime has been analyzed 
 in superspace Batalin-Vilkovisky (BV) formalism \cite{229p1}. 
Batalin-Vilkovisky (BV) formalism in the context of both 
the extended BRST, and the extended anti-BRST symmetries for Faddeev-Popov ghosts \cite{12a}-\cite{13a} along  
 a superspace formalism for it is also well understood \cite{14a}-\cite{15a}.
In this paper we will generalize the results that were
 obtained for the extended  BRST 
invariance of the perturbative 
quantum gravity  to include invariance
 under  extended anti-BRST symmetry also. We shall also 
 include noncommutativity of the spacetime coordinates, apart 
from the previously analyzed non-anticommutativity.

\section{Deformed Supermanifolds  }
In this section we will analyze noncommutative and non-anticommutative 
 supermanifolds. 
An elegant way to analyze perturbative quantum gravity with 
both noncommutative and non-anticommutative coordinates is in 
the language of supermanifolds~\cite{DeWitt}-\cite{Buchbinder}. 
The coordinates of the supermanifolds  can be written as 
\begin{equation}
\label{supercoordinates}
z  ^a  =x^a  +  y^a  ,
\end{equation}
where $x^a  $ are the bosonic coordinates with even Grassmann parity,
\begin{equation}
[x^a  ,x^b  ]=0,
\end{equation}
and $y^a  $ are  the fermionic coordinates with 
odd Grassmann parity,
\begin{equation}
\{y^a  ,y^b  \}\equiv y^a  y^b  +y^b  y^a  =0.
\end{equation}
So the metric can be now written as 
\begin{equation}
 ds^2 = g_{a  b  } dz  ^a   dz  ^b  .
\end{equation}
We want to impose noncommutativity and non-anticommutativity 
in such a way that the theory reduces to the 
usual noncommutative theory when there is no non-anticommutativity 
and it also reduces the usual non-anticommutative theory when there is no 
noncommutativity. This can be done by imposing the following 
relations
\begin{eqnarray}
[{\hat z  }^a  ,{\hat z  }^b  ]&
=& 2y^a  y^b  +i\theta^{a  b  }+O(\theta^2),
\nonumber \\ 
\{{\hat z  }^a  ,{\hat z  }^b  \}&=&2x^a  
x^b  +2i (x^a  y^b  +x^b  y^a  ) -\tau^{a  b  }+O(\tau^2).
\end{eqnarray}
Now in the limit $y^a  \rightarrow 0$ and
$\tau^{a  b  }\rightarrow 0$, we get 
\begin{eqnarray} [{\hat x}^a  ,{\hat
x}^b  ]&=&i\theta^{a  b  },
\nonumber \\
 \{{\hat x}^a  ,{\hat
x}^b  \}&=&2x^a   x^b  
\end{eqnarray}
and in the limit $x^a  \rightarrow 0$ and $\theta^{a  b  }\rightarrow 0$,
 we get 
\begin{eqnarray}
[{\hat y}^a  ,{\hat y}^b  ]&=&2y^a  y^b  ,
\nonumber \\ 
\{{\hat y}^a  ,{\hat y}^b  \}&=&\tau^{a  b  }.
\end{eqnarray}
We use Weyl
ordering and  express the Fourier transformation of this metric  as, 
\begin{equation}
\hat{g}^{(f)}_{ab} (\hat{z  }) =
\int d^4 k \pi e^{-i k \hat{z  }  } \;
g^{(f)}_{ab}  (k).
\end{equation}
Now  we  have a one to one map between a function of
$\hat{z  }$ to a function of ordinary
coordinates $ z  $ via
\begin{equation}
g^{(f)}_{ab} (z  )  =
\int d^4 k \pi e^{-i k x } \;
g^{(f)}_{ab} (k).
\end{equation}
\begin{equation}
g^{(f)ab}(z  )\circ    g^{(f)}_{ab}(z  ) =   \exp  (
\omega^{ab} \partial _a^2 \partial _b^1 )
 g^{(f)ab} (z  _1) g^{(f)}_{ab}  (z  _2)
\left. \right|_{z  _1=z  _2=z  }.
\end{equation}
where $\omega^{a  b  }$ is a nonsymmetric tensor
\begin{equation} 
\omega^{a  b  }=\tau^{a  b  }+\theta^{a  b  }.
\end{equation}
 Now  $R^{(f)a}_{bcd}$ given as,
\begin{equation}
R^{(f)a}_{bcd}=-\partial _d \Gamma^a_{bc} 
+\partial _c\Gamma^a_{bd}+\Gamma^a_{e c}\circ   \Gamma^e_{bd}
-\Gamma^a_{ed}\circ   \Gamma^e_{bc},
\end{equation}
and we also get  $R_{bc}=R^d_{bcd}$.
Thus, finally  $R^{(f)}$ is given by  
\begin{equation}
 R^{(f)} = g^{(f)ab}\circ    R_{ab}^{(f)}.
\end{equation}
 The Lagrangian density for pure gravity on the  supermanifolds  with 
 cosmological constant $\lambda$ can now be written as, 
\begin{equation}
 \mathcal{L}_c  = \sqrt{g}^{(f)}\circ   (R^{(f)} - 2\lambda),
\end{equation}
where we have adopted  units, such that  $16 \pi G = 1 $.

\section{Perturbative Quantum Gravity in BV Formulation}
In perturbative gravity on flat spacetime one splits the full 
metric $g_{ab}^{(f)}(z  )$ into $\eta_{ab}(z  )$ 
which is the metric for the
 background flat spacetime  and  $h_{ab} (z  )$ which is a small 
perturbation around the
 background spacetime. 
\begin{equation}
 g_{ab}^{(f)}(z  ) = \eta_{ab}(z  ) + h_{ab}(z  ).
\end{equation}
Here both $\eta_{ab}$ and $h_{ab}$ are defined on a supermanifold. 
The covariant derivatives along with the lowering and raising
 of indices are compatible with the 
metric for the background spacetime. The small 
perturbation $h_{ab}$, is viewed as the field that is to be quantized.

All the degrees of freedom in $h_{ab}$ 
are not physical as the  Lagrangian density for $h_{ab}$
 is invariant under the following 
 gauge transformations, 
\begin{eqnarray}
\delta_\Lambda h_{ab} &=&  D^e_{ab}\circ  
  \Lambda_e \nonumber \\ &=&
 [ \delta^e_b \partial _a  + \delta^e_a \partial  _b  
+ g^{ce} \circ    (\partial _c h_{ab})
 + \nonumber \\ &&g^{ec} \circ   
 h_{ac}\partial  _b  +\eta^{ec}\circ    h_{cb} \partial  _a ]
\circ    \Lambda_e. \label{eq}
\end{eqnarray}
In order to remove these unphysical  
degrees of freedom, we need to fix a gauge 
by adding a gauge-fixing term along with a ghost term. 
In Landau gauge 
 the sum of the gauge-fixing term and the ghost term can be expressed as
 \cite{gg}
\begin{eqnarray}
\mathcal{L}_g &=&-\frac{i}{2}s \overline{s} (h^{ab}\circ    h_{ab})
\end{eqnarray}
Now the sum of the ghost term, the gauge fixing term and the  
original classical Lagrangian density is invariant under the 
following the BRST
transformations
\begin{eqnarray}
s \,h_{ab} = D^e_{ab}\circ    c_e,  &&
s \,c^a - c_b \circ    \partial ^b c^a, \nonumber \\
s \,\overline{c}^a = - b^a, &&
s \,b^a =0, 
\end{eqnarray}
and the following anti-BRST transformations 
\begin{eqnarray}
\overline{s} \,h_{ab}= D^e_{ab}\circ    c_e, &&
\overline{s} \,c^a = b^a -  c^b
 \circ    \partial_b\overline{c}^a, \nonumber \\
\overline{s} \,\overline{c}^a = - \overline{c}^b 
\circ    \partial_b \overline{c}^a,&& 
\overline{s} \,b^a =  b^b\circ \partial_b \overline{c}^a,
\end{eqnarray}
where 
\begin{equation}
  D^e_{ab}= \delta^e_b \partial _a  + \delta^e_a \partial  _b  
+ g^{ce} \circ    (\partial _c h_{ab}) 
+ g^{ec} \circ    h_{ac}\partial  _b  +\eta^{ec}\circ    h_{cb}
 \partial  _a.
\label{eq}
\end{equation}
BV-formalism  is used to analyze
 the extended BRST and anti-BRST symmetries which is obtained by 
 first shifting  all the original fields as follows, 
\begin{eqnarray}
 h_{ab} &\to& h_{ab} -\tilde{h}_{ab},\nonumber \\ 
 c^a &\to& c^a -\tilde{c}^a, \nonumber \\ 
\overline{c}^a &\to& \overline{c}^a -\tilde{ \overline{c}}^a,\nonumber \\
b^a &\to& b^a -\tilde{b}^a,
\end{eqnarray}
and then requiring  the resultant  theory to be invariant 
under  the original BRST and anti-BRST transformations along with these
 shift transformations. 
This can be  achieved by letting the original
 fields to transform under the BRST transformations  as 
\begin{eqnarray}
s \,h_{ab} =\psi_{ab}, &&
s \,c^a = \phi^a , \nonumber \\
s \,\overline{c}^a  =\overline{\phi}^a, &&
s \,b^a =\rho^a,
\end{eqnarray}
and the shifted fields to transform under BRST transformations as 
\begin{eqnarray}
s \,\tilde{h}_{ab} =\psi_{ab}- {D^e_{ab}}'\circ   c'_e, &&
s \,\tilde{c}^a =\phi^a+  c_b'\circ    \partial^b c'_a, \nonumber \\
s \,\tilde{\overline{c}}^a =\overline{\phi}^a + {b^a}', &&
s \,\tilde{b}^a =\rho^a,
\end{eqnarray}
where 
\begin{eqnarray}
h'_{ab} = h_{ab}- \tilde{h}_{ab},  &&
c'^a = c^a - \tilde{c}^a , \nonumber \\
\overline{c}'^a  =\overline{c}^a - \overline{\tilde{c}}^a, &&
b'^a =b^a - \tilde{b}^a.
\end{eqnarray}
Here $\psi_{ab}, \phi^a, \overline{\phi}^a, \rho_a$
 are ghosts associated with the shift symmetry.
 Their BRST transformations vanishes,
\begin{eqnarray}
  s\, \psi_{ab} = 0, && s\, \phi^a =0 \nonumber \\ 
  s\,\overline{\phi}^a = 0, && s\,  \rho^a = 0.
\end{eqnarray}
We  define antifields, with opposite parity 
corresponding to all the original fields.
 These antifields have  following BRST transformations, 
\begin{eqnarray}
s \,h^*_{ab} =n _{ab}   , &&
s \,c^{*a} = m^{a}     , \nonumber \\
s \,\overline{c}^{*a}  =\overline{m}^{a}    , &&
s \,b^{*a} =\overline{n}^{a}.
\end{eqnarray}
The  BRST transformations of these Nakanishi-Lautrup fields vanishes,  
\begin{eqnarray}
s\, n _{ab}    = 0, && s\, m^{a}     = 0, \nonumber \\ s\,
 \overline{m}^{a}     = 0, && s\, \overline{n}^a =0.
\end{eqnarray}
We also let the original fields to transform 
under anti-BRST transformations as 
\begin{eqnarray}
\overline{s} \,h_{ab} &=&h^*_{ab} +
 {D_{ab}^e}' \circ    c_e', \nonumber \\
\overline{s} \,c^a &=& c^{*a} + {b^a }' - 2{c^b} '
\circ    \partial_b {\overline{c}^a} ' , \nonumber \\
\overline{s} \,\overline{c}^a  &=& \overline{c}^{*a} -
 {\overline{c}^b}'\circ    \partial_b{\overline{c}^a }',
 \nonumber \\ 
\overline{s} \,b^a &=& b^{*a} + 2{ b^b} ' \circ    \partial_b 
{\overline{c}^a }',
\end{eqnarray}
and the shifted fields to transform under the anti-BRST
 transformations as 
\begin{eqnarray}
\overline{s} \,\tilde{h}_{ab}  =h^*_{ab}, &&
\overline{s} \,\tilde{c}^a = c^{*a}, \nonumber \\
\overline{s} \,\tilde{\overline{c}}^a = \overline{c}^{*a},&&
\overline{s} \,\tilde{b}^a  =b^{*a}.
\end{eqnarray}
The anti-BRST transformations of the ghost fields is given by 
 \begin{eqnarray}
\overline{s} \,\psi_{ab}  &=&n _{ab}    + 2{D^e_{ab}}'\circ    
b_e '  - 2 {D_{ab}^e}'\circ     {c^b}' \circ    
\partial_b {\overline{c}^e }', \nonumber \\
\overline{s} \,\phi^a &=& m^{a}     - 2{ b^b}' \circ    
\partial_b {c^a }' 
+2 {\overline{c}^b}' \circ    \partial_b {c^c }' 
\circ    \partial_c {c^a}' , \nonumber \\
\overline{s} \,\overline{\phi}^a &=& \overline{n}^{a} 
-2 {b^b}'\circ    \partial_b { \overline{c}^a}' , \nonumber \\ 
\overline{s} \,\tilde{\rho}^a  &=&{m^{a}    }'.
\end{eqnarray}
The anti-BRST transformations of the antifields 
and the Nakanishi-Lautrup fields vanishes
\begin{eqnarray}
  {\overline{s}\, n _{ab}   } = 0, &&{ \overline{s}\, m^{a}    } = 0, 
\nonumber \\ \overline{s}\, \overline{m}^{a}     =0, && 
 \overline{s}{\overline{n}^{a}     } =0.
\end{eqnarray}
 Furthermore, the anti-BRST transformation 
of the Nakanishi-Lautrup fields also vanishes
\begin{eqnarray}
\overline{s}\, h^*_{ab} = 0, && 
 \overline{s} \,   c^{*a} = 0,
\nonumber \\ 
 \overline{s}\, \overline{c}^{*a} =0,&&
 \overline{s}\,  b^{*a}=0.
\end{eqnarray}

\section{Extended Superspace Formulation} 
In this section we will express the sum of 
the gauge fixing term and the ghost term in
 superspace formalism by using two anti-commutating  variable namely  
 $\theta$ and $\overline{\theta}$. Now  
we can define the following superfields, 
\begin{eqnarray}
 \omega_{ab} &=& h_{ab} + \theta \psi_{ab} + 
\overline{\theta} (h^*_{ab} + {D^e_{ab}}'\circ \overline{c}_e' )
\nonumber \\ &&  + 
  \theta \overline{\theta} (n _{ab}    + {D_{ab}^e}' \circ b_e  - 
{D^e_{ab}}' \circ {c^c}'\circ \partial_c
 \overline{c}_e'), \nonumber \\
 \tilde{\omega}_{ab} &=& \tilde{h}_{ab} +
 \theta (\psi_{ab} - {D_{ab}^e}' \circ c_e')
 + \overline{\theta} h_{ab}^* + 
\theta \overline{\theta} n _{ab}   ,\nonumber \\
\eta_a &=& c_a +\theta \phi_a + 
\overline{\theta}( c^{*}_a +{b_a}' - 2 {b^b}'\circ
 \partial_b {\overline{c}_a}'  ) \nonumber \\ && + 
  \theta \overline{\theta}
 (m_{a}     -2 {b^b}' \circ
 \partial_b {c_a}'
 +  {c^b}' \circ \partial_b
 {\overline{c}^c}' \circ
\partial_c {\overline{c}_a}'),
 \nonumber \\ 
\tilde{\eta}_a &=& \tilde{c}_a +
 \theta(\phi_a +c_b' 
 \partial^b c_a ) + 
\overline{\theta} c^{*}_a +
 \theta \overline{\theta} m_{a}    , \nonumber \\
\overline{\eta}_a &=& c_a 
+\theta \overline{\phi}_a 
+ \overline{\theta} ( \overline{c}^{*}_a
 - {\overline{c}^b}'\circ \partial_b 
{\overline{c}_a}'  ) \nonumber \\ &&
 \theta \overline{\theta} (\overline{m}_{a}     -
 2 {b^b}'\circ
 \partial_b {\overline{c}_a}' ) 
 , \nonumber \\ 
\tilde{\overline{\eta}}_a &=& \tilde{\overline{c}}_a +
 \theta(\overline{\phi}_a + b_a' )
 + \overline{\theta} \overline{c}^{*}_a + 
\theta \overline{\theta} \overline{m}_{a}    , \nonumber \\
\end{eqnarray}
Thus  we have 
\begin{eqnarray}
 \frac{\partial^2 }{ \partial \overline{\theta} \partial \theta}  
\tilde{\omega}^{*ab}\circ \tilde{\omega}_{ab} 
&=& - n _{ab}   \circ \tilde{h}^{ab}   -h^{*ab} (\psi_{ab} - {D^e_{ab}}' 
\circ c_e'), \nonumber \\
 \frac{\partial^2 }{ \partial \overline{\theta} \partial \theta}\tilde{\overline{\eta}}^*_a \circ
 \tilde{\eta}^a  &=&  -\overline{m}^{a}     \tilde{c}_a
 + \overline{c}^{*a} \circ (\psi_a  
  +  c_b' \circ \partial^b c_a' ),\nonumber \\
 -\frac{\partial^2 }{ \partial\overline{\theta}\partial \theta}\tilde{\overline{\eta}}^a \circ 
\tilde{\eta}^*_a  &=& m^{a}     \tilde{\overline{c}}_a  - 
c^*_a \circ ( \overline{\phi}^a  + {b^a}' ), \nonumber \\
- \frac{\partial^2 }{ \partial \overline{\theta} \partial\theta}  \tilde{f}^*_a \circ \tilde{f}^a 
 &=& B'^a \circ \tilde{b}_a + b^{*a}\circ  \rho_a.
\end{eqnarray}
Now we can express
 $\tilde{\mathcal{L}}_g $  as,
\begin{eqnarray}
 \tilde{\mathcal{L}}_g &=& 
\frac{\partial^2 }{ \partial \overline{\theta} \partial \theta} 
(\tilde{\omega}^{*ab}\circ \tilde{\omega}_{ab}
 +\tilde{\overline{\eta}}^*_a \circ \tilde{\eta}^a - 
\tilde{\overline{\eta}}^a \circ \tilde{\eta}^*_a -
 \tilde{f}^*_a \circ\tilde{f}^a ).
\end{eqnarray}

Furthermore,  we define $\Psi$ as, 
\begin{eqnarray}
 \Phi &=& \Psi + \theta s \Psi + \overline{\theta} \overline{s} \Psi +
\theta \overline{\theta} s \overline{s} \Psi.  
\end{eqnarray}
Thus, we can express $\mathcal{L}_g$   as,
\begin{equation}
 \mathcal{L}_g =  \frac{\partial^2  \Phi}{\partial \overline{\theta}\partial
 \theta }.
\end{equation}
Now the complete Lagrangian in the superspace formalism is given by 
 \begin{eqnarray}
  \mathcal{L} &=&  \frac{\partial^2  \Phi}{\partial\overline{\theta} \partial\theta }
  + \frac{\partial^2   }{\partial\overline{\theta}\partial \theta }
(\tilde{\omega}^*_{ab}\circ\tilde{\omega}^{ab} + 
\tilde{\overline{\eta}}^*_a \circ\tilde{\eta}^a-
\tilde{\overline{\eta}}^a \circ
\tilde{\eta}_a^* -
\tilde{\overline{\eta}}^a\circ \tilde{\eta}_a^* -
 \tilde{f}^*_a\circ \tilde{f}^a) \nonumber \\ && 
+\mathcal{L}_c( h_{ab} -\tilde{h}_{ab}).
 \end{eqnarray}
This Lagrangian density is manifestly
 invariant under the extended BRST transformations. 
Furthermore, it is also invariant under on-shell extended anti-BRST 
transformations. 
\section{Conclusion}
In this paper we have  analyzed 
 perturbative quantum gravity  with 
 noncommutative and non-anticommutative coordinates
on  supermanifolds. After analyzing the BRST and the anti-BRST 
symmetries of this theory, 
we analyzed the extended BRST and the anti-BRST symmetries associated with 
it. Extended BRST and extended anti-BRST transformations  were obtained by 
requiring the theory to be invariant under the original BRST and the 
original anti-BRST transformations along with the shift transformations. 
This was done by the using the BV-formulism. Then these extended BRST and
anti-BRST symmetries were elegantly written down using  extended 
superspace formulism.

The supermanifold formulism is ideally suited to 
study supersymmetric theories. So it will be interesting to 
perform a similar analyses on   supergravity theories. 
It will also be interesting to generalize this work to anti-de Sitter 
spacetime as then we will be able to analyze the superconformal theories 
dual to this model of quantum gravity. The generalization of this work to 
anti-de Sitter spacetime can be easily done. This is because   there are 
no infrared divergences in the ghosts in this spacetime. However, 
the generalization of this work to de Sitter spacetime will not be trivial, 
due to the infrared divergences in de Sitter spacetime \cite{fa}. However, 
if this problem can some how be resolved then we can possible 
analyze the implications of this model on inflation.

\end{document}